\documentclass[10pt]{article}
\usepackage[a4paper]{geometry}
\usepackage[utf8]{inputenc}
\usepackage[T1]{fontenc}
\usepackage{setspace,authblk}
\usepackage{amsmath,mathrsfs,amsfonts,amsthm,amssymb}

\usepackage{tikz}

\usepackage[english]{babel}

\usepackage{graphicx}
 
\usepackage{hyperref}

\title{\singlespace {A representation theorem for events within lattice structures of state-spaces}}

\author[]{Alex A.T. Rathke\thanks{NECCT/FEA-RP/USP, University of S\~ao Paulo. \texttt{alex.rathke@alumni.usp.br}}}

\date{\today}

\theoremstyle{plain}
\newtheorem{theorem}{Theorem}
\newtheorem*{theorem*}{Theorem}

\newtheorem*{proposition*}{Proposition}

\newtheorem*{remark*}{Remark}

\newtheorem*{condition*}{Condition}

\newtheorem*{definition*}{Definition}

\newtheorem*{assumption*}{Assumption}

\begin{document}

\maketitle

\begin{abstract} 

For the standard lattice model of information structures, we derive a reduced poset representation which provides the same informational content as the complete lattice structure which derives it. Rational agents can recover the complete lattice of events by means of the reduced poset alone. We find that both structures provide isomorphic models under mild conditions.

\end{abstract}


\section{Introduction} \label{Introduction}

We introduce the lattice model of information structures of agents proposed by \cite{fukuda2023,fukuda2024,heifetz2006,heifetz2013,schipper2021}. Let a lattice $(\mathcal{S}, \succeq)$ of disjoint state-spaces $S \in \mathcal{S}$ be partially ordered by the level of "expressiveness" of each space, where $S^{\prime} \succeq S$ indicates that $S^{\prime}$ is at least as expressive as $S$. The order $S^{\prime} \succeq S$ means that states $\omega \in S^{\prime}$ may describe states in $S$ in a richer or more expressive way \cite{fukuda2023,fukuda2024,heifetz2006,heifetz2013,schipper2021}. The set $\Omega := \bigcup_{\mathcal{S}} (S)$ defines the union of all states across all state-spaces $S \in \mathcal{S}$. For ease of exposition, we assume a full sigma-algebra $2^{S}$ for every state-space $S \in \mathcal{S}$, unless explicitly referred otherwise.

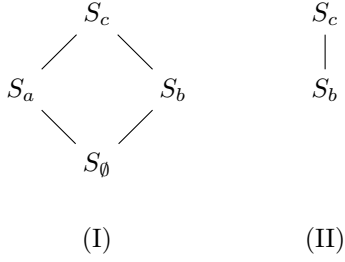
\begin{figure}[htb!]
\begin{center}
\begin{tikzpicture}
	\node (max1) at (1,3) {$ S_{c}$};
	\node (a1) at (0,2) {$ S_{a}$};
	\node (b1) at (2,2) {$ S_{b}$};
	\node (min1) at (1,1) {$ S_{\emptyset}$};
	\node (mmin1) at (1,0) {(I)};
	\draw (min1) -- (a1) -- (max1);
	\draw (min1) -- (b1) -- (max1);
	\node (max2) at (4,3) {$ S_{c}$};
	\node (a2) at (4,2) {$ S_{b} $};
	\node (mmin2) at (4,0) {(II)};
	\draw (a2) -- (max2);
\end{tikzpicture}
\end{center}
\caption{(I) Hasse diagram for the lattice $(\mathcal{S}, \succeq)$, with $\mathcal{S} = \{S_{a}, S_{b}, S_{c} \}, S_{c} \succeq S_{a}, S_{c} \succeq S_{b} $, the empty state-space $S_{\emptyset}$ included. (II) Hasse diagram for the filter $(\mathcal{S}_{S_{b}}, \succeq)$, $\mathcal{S}_{S_{b}} = \{ S_{b}, S_{c} \}, S_{c} \succeq S_{b}$.} \label{hd1}
\end{figure}

Refer to the \emph{lattice filter} equal to $(\mathcal{S}_{S}, \succeq), \mathcal{S}_{S} :\subseteq \mathcal{S}$ as the non-empty sub-lattice within $( \mathcal{S}, \succeq)$ with respect to the \emph{meet} equal to $\inf (\mathcal{S}_{S}) = S$, also called the filter base, satisfying the usual conditions, i.e. for every $S^{\prime}, S^{\prime \prime} \in \mathcal{S}_{S}$, we have $S^{\prime} \succeq S$ and $S^{\prime \prime} \succeq S$, and for every $S^{\prime} \in \mathcal{S}_{S}$ and $S^{\prime \prime} \in \mathcal{S}$, condition $S^{\prime \prime} \succeq S^{\prime}$ implies $S^{\prime \prime} \in \mathcal{S}_{S}$ \cite{gratzer2011,davey2002}. For example, the lattice\footnote{As a notation device, our study regards that all sets of state-spaces $\mathcal{S}$ include the empty state-space $S_{\emptyset} = \emptyset$, so our example regards the set of state-spaces defined as $\mathcal{S} = \{ S_{\emptyset}, S_{a}, S_{b}, S_{c} \} = \{S_{a}, S_{b}, S_{c} \}$.} with $\mathcal{S} = \{S_{a}, S_{b}, S_{c} \}$, orders $S_{c} \succeq S_{a}, S_{c} \succeq S_{b} $ derives e.g. the filter $(\mathcal{S}_{S_{b}}, \succeq)$ with $\mathcal{S}_{S_{b}} = \{ S_{b}, S_{c} \}, S_{c} \succeq S_{b}$, base $\inf (\mathcal{S}_{S_{b}}) = S_{b}$, see diagrams in Fig. \ref{hd1}. The \emph{join} of a filter is equal to $\sup ( \mathcal{S}_{S} ) = \sup ( \mathcal{S} )$, for all $S \in \mathcal{S}$ \cite{gratzer2011,davey2002}.

Following the studies of \cite{fukuda2024,heifetz2006,heifetz2013,schipper2021}, define a surjective projection from the more expressive to the less expressive state-spaces equal to $r^{S^{\prime}}_{S} : S^{\prime} \rightarrow S$, for each order $S^{\prime} \succeq S$, the projective condition $r^{S}_{S} (\omega) = \omega$ for any states $\omega \in S$, all spaces $S \in \mathcal{S}$. The state $r^{S^{\prime}}_{S} (\omega^{\prime}) \in S$ is the representation of the state $\omega^{\prime} \in S^{\prime}$ within the less expressive state-space $S$ \cite{fukuda2023,fukuda2024,heifetz2006,heifetz2013,schipper2021}. Assume that the projection $r^{S^{\prime}}_{S}$ satisfies composition, for $S^{\prime \prime} \succeq S^{\prime } \succeq S$ implies $r^{S^{\prime \prime}}_{S} = r^{S^{\prime}}_{S}  \circ  r^{S^{\prime \prime}}_{S^{\prime}}$. The inverse projection $(r^{S^{\prime}}_{S})^{-1} (\omega) \subseteq S^{\prime} $ refers to the set of states in the more expressive state-space $S^{\prime}$ from which the state $\omega \in S$ is projected\footnote{For sets of states, the condition $r^{S^{\prime}}_{S} ( \{ \omega^{\prime}, \omega^{\prime \prime} \}) = \{ r^{S^{\prime}}_{S} (\omega^{\prime}) \} \cup \{ r^{S^{\prime}}_{S} (\omega^{\prime \prime}) \}$ derives directly, with $\omega^{\prime}, \omega^{\prime \prime} \in S^{\prime}$. It implies $\bigcup_{S} (r^{S^{\prime}}_{S})^{-1} (\omega) = (r^{S^{\prime}}_{S})^{-1} (S) \subseteq S^{\prime}$, all $\omega \in S$.}.

\begin{figure}[htb!]
\begin{center}
\begin{tikzpicture} 
	\node (max0) at (0.25,3.3) {$ S_{c} = \{ $};
	\node (max1) at (1,3.3) {$ c_{1} ,$};
	\node (max2) at (1.5,3.3) {$ c_{2} ,$};
	\node (max3) at (2.05,3.3) {$ c_{3} \} $};
	\node (max1s) at (0.9,3.1) { };
	\node (max2s) at (1.4,3.1) { };
	\node (max3s) at (2,3.14) { };
	\node (a0) at (-2.05,1.05) {$ S_{a} = \{ $};
	\node (a1) at (-1.4,1) {$ a ,$};
	\node (a2) at (-0.85,1.05) {$ \neg a \} $};
	\node (a3) at (-1,1.04) { };
	\node (b0) at (3.05,1) {$ S_{b} = \{ $};
	\node (b1) at (3.7,1) {$ b , $};
	\node (b2) at (4.3,1) {$ \neg b \} $};
	\node (min0) at (0.7,-1.3) {$ S_{\emptyset} =$};
	\node (min1) at (1.3,-1.3) {$ \emptyset$};
	\draw [dotted, out=90, in=230] (a1) to (max1s);
	\draw [dotted, out=80, in=230] (a1) to (max2s);
	\draw [dotted, out=70, in=230] (a3) to (max3s);
	\draw [dotted, out=110, in=270] (b1) to (max1s);
	\draw [dotted, out=100, in=270] (b2) to (max2s);
	\draw [dotted, out=90, in=270] (b2) to (max3s);
	\draw [dotted, out=270, in=110] (a1) to (min1);
	\draw [dotted, out=270, in=110] (a2) to (min1);
	\draw [dotted, out=270, in=70] (b1) to (min1);
	\draw [dotted, out=270, in=70] (b2) to (min1);
\end{tikzpicture}
\end{center}
\caption{Refer to the lattice example in Fig. \ref{hd1}, (I), edges omitted. Diagram for the projections in dotted lines: $r^{S_{c}}_{S_{a}} (c_{1}) = r^{S_{c}}_{S_{a}} (c_{2}) = a$, $r^{S_{c}}_{S_{a}} (c_{3}) = \neg a$, $r^{S_{c}}_{S_{b}} (c_{1}) = b$, $ r^{S_{c}}_{S_{b}} (c_{2}) = r^{S_{c}}_{S_{b}} (c_{3}) = \neg b$.} \label{hdd}
\end{figure}
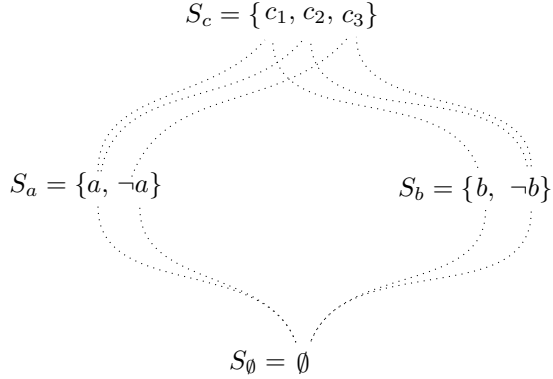

For example, assume the same lattice $(\mathcal{S},\succeq)$ in Fig \ref{hd1} with state-spaces $S_{a} = \{ a, \neg a \}, S_{b} = \{ b, \neg b \}, S_{c} = \{ c_{1}, c_{2}, c_{3} \} $. Assume the projections $r^{S_{c}}_{S_{a}} (c_{1}) = r^{S_{c}}_{S_{a}} (c_{2}) = a$, $r^{S_{c}}_{S_{a}} (c_{3}) = \neg a$, $r^{S_{c}}_{S_{b}} (c_{1}) = b$, $ r^{S_{c}}_{S_{b}} (c_{2}) = r^{S_{c}}_{S_{b}} (c_{3}) = \neg b$. We have e.g. $(r^{S_{c}}_{S_{a}})^{-1} (a) = \{ c_{1}, c_{2} \}, (r^{S_{c}}_{S_{a}})^{-1} (\neg a) = \{ c_{3} \}$. The projections within the lattice structure are presented in dotted lines in Fig. \ref{hdd}.

In the lattice model, all operations on subsets take into account the disjoint state-spaces $S \in \mathcal{S}$ in which they are included. Projections $r^{S^{\prime}}_{S}$ apply to define events\footnote{Two comments apply. First, events in $S$ are formally well-defined with respect to the elements of its sigma algebra $2^{S}$. We follow the unambiguously common notation so to refer to events with respect to $S$. Second, the standard definition of events as simple subsets of the full union space $\Omega$ does not model the underlying order of expressiveness of state-spaces as intended by the lattice model.} regarding subsets of the full union space $\Omega$ while preserving information about the order of expressiveness across the state-spaces in which they are defined. For any state $\omega \in S$, each space $S \in \mathcal{S}$, the set of states projected to $\omega$ within the filtered spaces in $\mathcal{S}_{S}$ defines the upper set equal to \cite{fukuda2024,heifetz2006,heifetz2013}

\begin{equation} \label{ip1}
\{ \omega \}^{\uparrow} := \bigcup_{\mathcal{S}_{S}} (r^{S^{\prime}}_{S})^{-1} (\omega) . \\
\end{equation}

It follows that for a set $E \subseteq S$, we have $E^{\uparrow} = \bigcup_{\mathcal{S}_{S}} (r^{S^{\prime}}_{S})^{-1} (E)$. In current studies, an event is defined as a pair $(E^{\uparrow}, S) \in 2^{\Omega} \times \mathcal{S}$, where $E^{\uparrow}$ is the upper set obtained in Eq. \ref{ip1}, and $S$ is the base-space satisfying $E \subseteq \sup ( \inf(\mathcal{S}_{S}) ) = S$, in shorthand $E \subseteq S$, each $S \in \mathcal{S}$ \cite{fukuda2024,heifetz2006,heifetz2013}, i.e. the base $S$ provides the least expressive state-space in which the event $(E^{\uparrow}, S)$ can be described. In Fig \ref{hd1} and \ref{hdd}, the upper sets e.g. $\{ a \}^{\uparrow}$, $\{ b \}^{\uparrow}$ derive the events equal to $(\{a, c_{1}, c_{2} \}, S_{a})$, $(\{b, c_{1} \}, S_{b})$ respectively.

Operations on events of the form $(E^{\uparrow}, S) \in 2^{\Omega} \times \mathcal{S}$ affect their base-spaces. The negation of an event refers to the complement with respect to its base, so defined as $\neg (E^{\uparrow}, S) := ((S \setminus E)^{\uparrow}, S)$ \cite{fukuda2023,fukuda2024,heifetz2006,heifetz2013,schipper2021}. The negation of a complete state-space as $E = S$ implying $\neg(S^{\uparrow},S) = ((S \setminus S)^{\uparrow}, S)$ derives the upper set $(S \setminus S)^{\uparrow} = \emptyset^{\uparrow} = \Omega$, see Eq. \ref{ip1}, for it derives the event $(\Omega, \emptyset)$, the base-space changes to $S_{\emptyset} = \emptyset$ for every $S \in \mathcal{S}$, which satisfies the least-expressiveness condition $\emptyset \subseteq \sup ( \inf \mathcal{S} ) = S_{\emptyset} = \emptyset$\footnote{By the definition of negation, we find $\neg(S^{\uparrow},S) = \neg \neg(S^{\uparrow},S) = \neg \neg \neg(S^{\uparrow},S) = \ldots$ for every $S \in \mathcal{S}$. Our study assumes one empty set satisfying $\emptyset = (S \setminus S) = (S^{\prime} \setminus S^{\prime})$ for all states $S, S^{\prime} \in \mathcal{S}$. In contrast, current studies demand the definition of a specific empty set for each state-space, so that $( S \setminus S ) \neq (S^{\prime} \setminus S^{\prime} )$, for each $S, S^{\prime} \in \mathcal{S}$, see \cite{fukuda2023,fukuda2024,heifetz2006,heifetz2013,schipper2021}. We remark that the condition $( S \setminus S ) \neq (S^{\prime} \setminus S^{\prime} )$ in current studies violates the axioms of separation and extensionality of the classical ZFC set theory. Besides, it implies that the range of the upper sets as defined in Eq. \ref{ip1} expands to $2^{\Omega} \cup (S \setminus S) \cup (S^{\prime} \setminus S^{\prime}) \cup \ldots$ regarding all state-spaces $S, S^{\prime} \in \mathcal{S}$.}. Otherwise, negation preserves base-spaces.

The intersection of events of the form $(E^{\uparrow}, S)$ is not the same as the standard set intersection. For the general case, we have events with base-spaces which are non-ordered, for their upper sets may eventually intersect at more expressive state-spaces. In our example with events $(\{ a \}^{\uparrow}, S_{a})$, $(\{b \}^{\uparrow}, S_{b})$, the intersection $\{a \}^{\uparrow} \cap \{ b \}^{\uparrow} = \{a, c_{1}, c_{2} \} \cap \{b, c_{1} \} = \{ c_{1} \} \subseteq S_{c} $ is defined within the more expressive state-space $S_{c} = \sup (S_{a}, S_{b}) = \sup ( \inf (\mathcal{S}_{S_{a}}), \inf (\mathcal{S}_{S_{b}}))$, which refers to the join of $S_{a},S_{b}$ \cite{gratzer2011,davey2002}, see Fig. \ref{hd1}. Hence, for any collection of $L$ events $( E^{\uparrow}_{i}, S_{i}), i \in L$, where $\sup_{L} (S_{i}) = \sup_{L} ( \inf ( \mathcal{S}_{S_{i}} ) )$ is the join of all their base-spaces $S_{i}$, the \emph{conjunction} of events is defined as \cite{fukuda2023,fukuda2024,heifetz2006,heifetz2013}

\begin{equation} \label{ip2}
\bigwedge_{i \in L} (E^{\uparrow}_{i}, S_{i}) := \left( \bigcap_{i \in L} (E^{\uparrow}_{i}) , \sup_{i \in L} (S_{i}) \right), \\
\end{equation}

\noindent which is an event with base-space $\sup_{L} (S_{i}) $. We find that an empty intersection is such that $\bigcap_{i \in L} (E^{\uparrow}_{i}) = \emptyset \rightarrow (\emptyset, \sup \mathcal{S} )$ is an event \cite{fukuda2023,fukuda2024,heifetz2006,heifetz2013}, i.e. remark the difference $\bigcap_{i \in L} (E^{\uparrow}_{i}) = \emptyset \neq \emptyset^{\uparrow} = \Omega$. In our example in Fig. \ref{hd1}, we find e.g. a non-empty conjunction equal to $(\{ a \}^{\uparrow}, S_{a}) \wedge (\{ b \}^{\uparrow}, S_{b}) = ( \{ c_{1} \}, S_{c} )$, base-space $\sup ( S_{a}, S_{b} ) = S_{c}$.

The \emph{disjunction} of $L$ events $( E^{\uparrow}_{i}, S_{i}), i \in L$ is usually defined as \cite{fukuda2023,fukuda2024,heifetz2006,heifetz2013}

\begin{equation} \label{ip3}
\bigvee_{i \in L} (E^{\uparrow}_{i}, S_{i}) := \neg \left( \bigwedge_{i \in L} \neg (E^{\uparrow}_{i}, S_{i}) \right) = \left[ \left( \sup_{i \in L} (S_{i}) \setminus \bigcap_{i \in L} ( (S_{i} \setminus E_{i})^{\uparrow} ) \right)^{\uparrow}, \sup_{i \in L} (S_{i}) \right] , \\
\end{equation}

\noindent which is an event with base-space $\sup_{L} (S_{i}) $. In our example in Fig. \ref{hd1}, Eq. \ref{ip3} provides e.g. a disjunction of events equal to $(\{ a \}^{\uparrow}, S_{a}) \vee (\{ b \}^{\uparrow}, S_{b}) = ( \{ c_{1},c_{2} \}, S_{c} )$, base-space $\sup ( S_{a}, S_{b} ) = S_{c}$.

Events of the form $(E^{\uparrow}, S) \in 2^{\Omega} \times \mathcal{S}$ induce a type of partial order which may be defined as follows. For any two events $(E^{\uparrow}, S), (F^{\uparrow}, S^{\prime})$, define a partial order as $\leqslant : (F^{\uparrow}, S^{\prime}) \leqslant (E^{\uparrow}, S)$ iff $F^{\uparrow} \subseteq E^{\uparrow}, S^{\prime} \succeq S$ \cite{fukuda2024,heifetz2013}. The partial order $\leqslant$ may be interpreted as a decreasing order of expressiveness of events. In our example in Fig. \ref{hd1} and \ref{hdd}, the order of events $(\{ c_{1} \}^{\uparrow}, S_{c}) \leqslant (\{b \}^{\uparrow}, S_{b})$ follows from the conditions $\{c_{1} \}^{\uparrow} = \{ c_{1} \} \subseteq \{b \}^{\uparrow} = \{b, c_{1} \}$, $S_{c} \succeq S_{b}$. In our interpretation, it means that the event $(\{b \}^{\uparrow}, S_{b})$ is the description of the event $(\{ c_{1} \}^{\uparrow}, S_{c})$ within the less expressive state-space $S_{b}$. For events with the same base-space $S$, the partial order $\leqslant$ refers to a standard set inclusion $\subseteq$, so following the usual interpretation, i.e. we may say that the event $(F^{\uparrow}, S)$ is finer than $(E^{\uparrow}, S)$ if we have $(F^{\uparrow}, S) \leqslant (E^{\uparrow}, S)$.

Let the collection of all events $(E^{\uparrow}, S) \in \mathcal{E}$ deriving from a general lattice of state-spaces $(\mathcal{S}, \succeq)$ be partially ordered with respect to $\leqslant$. We obtain a new lattice of events $(\mathcal{E}, \leqslant)$ bounded by the meet $\inf \mathcal{E} = (\emptyset, \sup \mathcal{S})$ and the join $\sup \mathcal{E} = (\Omega, \emptyset)$ \cite{fukuda2023,fukuda2024,heifetz2013}. Pairwise meets within the lattice $(\mathcal{E},\leqslant)$ satisfy the conjunction of events in Eq. \ref{ip2}, following the definition of $\leqslant$. On the other hand, we remark that pairwise joins within the lattice $(\mathcal{E},\leqslant)$ do not satisfy the disjunction defined in Eq. \ref{ip3}\footnote{Remark that pairwise joins of events in $(\mathcal{E}, \leqslant)$ occur at the less expressive base-spaces, following the definition of the partial order $\leqslant$.}. The lattice of events $(\mathcal{E}, \leqslant)$ orders the descriptions of all events in $\mathcal{E}$ with respect to the level of expressiveness of their base-spaces.

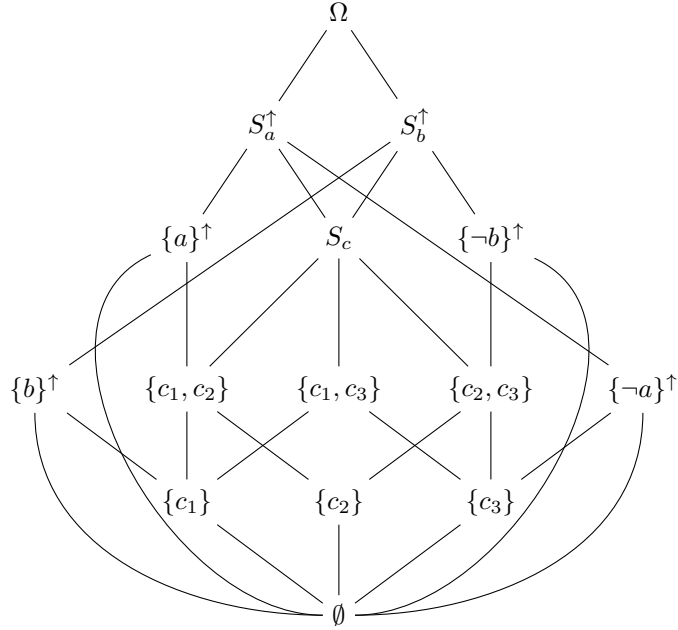
\begin{figure}[htb!]
\begin{center}
\begin{tikzpicture}
	\node (min) at (0,1) {$ \emptyset$};
	\node (c1) at (-2,2.5) {$ \{ c_{1} \}$};
	\node (c2) at (0,2.5) {$ \{ c_{2} \}$};
	\node (c3) at (2,2.5) {$ \{ c_{3} \}$};
	\node (cc1) at (-2,4) {$ \{ c_{1}, c_{2} \}$};
	\node (cc2) at (0,4) {$ \{ c_{1}, c_{3} \}$};
	\node (cc3) at (2,4) {$ \{ c_{2}, c_{3} \}$};
	\node (ccc) at (0,6) {$ S_{c}$};
	\node (b) at (-4,4) {$ \{ b \}^{\uparrow}$};
	\node (na) at (4,4) {$ \{ \neg a \}^{\uparrow}$};
	\node (a) at (-2,6) {$ \{ a \}^{\uparrow} $};
	\node (nb) at (2,6) {$ \{ \neg b \}^{\uparrow} $};
	\node (aaa) at (-1,7.5) {$ S_{a}^{\uparrow} $};
	\node (bbb) at (1,7.5) {$ S_{b}^{\uparrow} $};
	\node (max) at (0,9) {$ \Omega $};
	\draw (min) -- (c1) -- (cc1) -- (ccc) -- (aaa) -- (max);
	\draw (min) -- (c2) -- (cc1);
	\draw (c2) -- (cc3);
	\draw (c1) -- (b) -- (bbb) -- (max);
	\draw (cc1) -- (a) -- (aaa);
	\draw (cc3) -- (nb) -- (bbb);
	\draw (c3) -- (na);
	\draw (c3) -- (cc2);
	\draw (c1) -- (cc2) -- (ccc) -- (bbb);
	\draw (na) -- (aaa);
	\draw (min) -- (c3) -- (cc3) -- (ccc);
\end{tikzpicture}
\end{center}
\caption{Hasse diagram for the lattice of events $(\mathcal{E}, \leqslant)$ deriving from the lattice of state-spaces $(\mathcal{S}, \succeq)$ in Fig. \ref{hd1}, (I), with $\mathcal{S} = \{S_{a}, S_{b}, S_{c} \}$, $S_{c} \succeq S_{a}, S_{c} \succeq S_{b} $, $S_{a} = \{ a, \neg a \}$, $S_{b} = \{ b, \neg b \}$, $S_{c} = \{ c_{1}, c_{2}, c_{3} \}$, the projections across state-spaces are presented in Fig. \ref{hdd}. We omit the base-spaces of events by applying a concise notation equal to $(E^{\uparrow},S) = E^{\uparrow}$ for all events $(E^{\uparrow},S) \in \mathcal{E}$, the least-expressiveness condition $E \subseteq S$ implying unambiguous base-spaces.} \label{hde}
\end{figure}

For example, assume the lattice of state-spaces $(\mathcal{S},\succeq)$ in Fig. \ref{hd1} and \ref{hdd}, $\mathcal{S} = \{S_{a}, S_{b}, S_{c} \}$, orders $S_{c} \succeq S_{a}, S_{c} \succeq S_{b} $, state-spaces $S_{a} = \{ a, \neg a \}, S_{b} = \{ b, \neg b \}, S_{c} = \{ c_{1}, c_{2}, c_{3} \} $, and assume the same projections as in Fig. \ref{hdd}. We obtain the lattice of events $(\mathcal{E}, \leqslant)$ ordered by decreasing expressiveness as presented in Fig. \ref{hde}. (For a more concise notation, we omit the base-spaces of events in the Hasse diagram in Fig. \ref{hde}, so to refer to each event by its upper set only, $(E^{\uparrow},S) = E^{\uparrow}$ for all events in $\mathcal{E}$, e.g. $( \{ a \}^{\uparrow}, S_{a}) = \{ a \}^{\uparrow}$, $(\{ c_{1}, c_{2} \}^{\uparrow}, S_{c}) = \{ c_{1}, c_{2} \}$, $( \{ b, \neg b \}^{\uparrow}, S_{b}) = (S_{b}^{\uparrow}, S_{b}) = S_{b}^{\uparrow}$, $(\emptyset, \sup \mathcal{S}) = \emptyset$, $(\Omega, \emptyset) = \Omega$. The least-expressiveness condition of base-spaces $E \subseteq S$ for all events $(E^{\uparrow},S) \in \mathcal{E}$ implies unambiguous bases\footnote{In detail, every upper set $E^{\uparrow} \in 2^{\Omega}$ is already defined with respect to its base-space $S \in \mathcal{S}$, see Eq. \ref{ip1}.} \cite{fukuda2023,fukuda2024,heifetz2006,heifetz2013}.)

\section{Modelling a reduced poset of events} \label{Modelling a reduced poset of events}

As the number of states $\omega \in S$ and state-spaces $S \in \mathcal{S}$ increases, the lattice of events $(\mathcal{E}, \leqslant)$ becomes noticeably more intricate when compared with the lattice of state-spaces $(\mathcal{S}, \succeq)$ which derives it. Theorem \ref{t01} regards a relation between both lattices which formalises our interpretation of the partial order $\leqslant$ as a decreasing order of expressiveness of events.

\begin{theorem} \label{t01}
There exists an order reversing $h^{\prime} : \mathcal{S} \rightarrow \mathcal{E}$ between the lattice of state-spaces $(\mathcal{S}, \succeq)$ and the lattice of events $(\mathcal{E}, \leqslant)$.
\begin{proof}
For the lattices $(\mathcal{S}, \succeq), (\mathcal{E}, \leqslant)$, a monotone order reversing $h^{\prime} : \mathcal{S} \rightarrow \mathcal{E}$ is defined as for all $S,S^{\prime} \in \mathcal{S}$, if $S^{\prime} \succeq S$, then $h^{\prime}(S^{\prime}) \leqslant h^{\prime}(S)$ \cite{gratzer2011,davey2002}. For the partial order $\succeq$, we have $S^{\prime} \succeq S \rightarrow \sup (S^{\prime},S) = S^{\prime}, \inf (S^{\prime},S) = S$. By the definition of the partial order $\leqslant$, we have $(F^{\uparrow}, S^{\prime}) \leqslant (E^{\uparrow},S) \rightarrow \sup((E^{\uparrow},S), (F^{\uparrow}, S^{\prime})) = (E^{\uparrow},S)$, $\inf((E^{\uparrow},S), (F^{\uparrow}, S^{\prime})) = (F^{\uparrow},S^{\prime})$. 
\end{proof}
\end{theorem}

Within a state-space $S \in \mathcal{S}$, the expressiveness of events $(E^{\uparrow},S) \in \mathcal{E}$ is consistent with the standard approach where events are the subsets of $S$ and the order of expressiveness is defined by set inclusion $\subseteq$, i.e. $(F^{\uparrow},S) \leqslant (E^{\uparrow},S)$ iff $F \subseteq E$ \cite{fukuda2023,heifetz2006,heifetz2013}. We find that for each state-space $S \in \mathcal{S}$, there exists an order embedding of the sigma-algebra $2^{S}$ into the collection of all events $\mathcal{E}$. It means that we may embed any standard model $(2^{S},\subseteq)$ into the lattice model $(\mathcal{E},\leqslant)$ while preserving the order of expressiveness of events.

\begin{theorem} \label{t02}
Regard the sigma-algebra $2^{S}$, every $S \in \mathcal{S}$. There exists an order embedding $f : 2^{S} \rightarrow \mathcal{E}$ of the lattice of set inclusion $(2^{S}, \subseteq)$ into the lattice of events $(\mathcal{E}, \leqslant)$, for each state-space $S \in \mathcal{S}$.
\begin{proof}
An order embedding $f : 2^{S} \rightarrow \mathcal{E}$ is defined as for all $E,F \in 2^{S}$, we have $F \subseteq E$ iff $f(F) \leqslant f(E)$ \cite{gratzer2011,davey2002}. By definition, the partial order $\leqslant$ refers to the standard set inclusion $\subseteq$ for events with the same base-space $S \in \mathcal{S}$.
\end{proof}
\end{theorem}

Theorem \ref{t01} shows that for each event $(F^{\uparrow}, \sup \mathcal{S}) \in \mathcal{E}$ with the most expressive base-space $\sup \mathcal{S}$, there is always an event $(E^{\uparrow}, S) \in \mathcal{E}$ satisfying $(F^{\uparrow}, \sup \mathcal{S}) \leqslant (E^{\uparrow}, S)$, meaning that the event $(E^{\uparrow}, S)$ is the description of the event $(F^{\uparrow}, \sup \mathcal{S})$ within the state-space $S$. For example, in Fig. \ref{hde}, e.g. the event $(\{c_{1} \}^{\uparrow}, S_{c})$ is described within the state-space $S_{b}$ by the event $(\{b \}^{\uparrow}, S_{b})$, following from the order $(\{c_{1} \}^{\uparrow}, S_{c}) \leqslant (\{b \}^{\uparrow}, S_{b})$, while the event $(\{c_{2} \}^{\uparrow}, S_{c})$ is described in $S_{c}$ only.

Theorem \ref{t02} shows that we may embed the lattice of set inclusion $(2^{(\sup \mathcal{S})}, \subseteq)$ of the sigma-algebra $2^{(\sup \mathcal{S})}$ of all subsets of the most expressive state-space $\sup \mathcal{S}$ into the full lattice of events $(\mathcal{E},\leqslant)$. In our example, $\sup \mathcal{S} = S_{c}$, we visualise in Fig. \ref{hde} that the diagram of the lattice $(2^{S_{c}}, \subseteq)$ is embedded into the full lattice of events $(\mathcal{E}, \leqslant)$, and the order of inclusion of subsets of $S_{c}$ is preserved.

Refer to $\mathcal{E}^{\prime} \subseteq \mathcal{E}$ as the collection of the finest least-expressive descriptions of the most expressive events $(E^{\uparrow}, \sup \mathcal{S}) \in \mathcal{E}$. In detail, we define $\mathcal{E}^{\prime} \subseteq \mathcal{E}$ as the collection of events satisfying

\begin{equation} \label{rp1}
\begin{array}{rl}
\mathcal{E}^{\prime} &:= \{ (E^{\uparrow},S) \in \mathcal{E} : ( (E^{\uparrow} \cap \sup \mathcal{S})^{\uparrow}, \sup \mathcal{S}) \leqslant ( (E^{\uparrow} \cap S^{\prime \prime})^{\uparrow}, S^{\prime \prime}) \leqslant \ldots \\
\\
&\ldots \leqslant ( (E^{\uparrow} \cap S^{\prime})^{\uparrow}, S^{\prime}) \leqslant (E^{\uparrow},S) \leqslant (F^{\uparrow}, S) \} \cup \{ (\emptyset, \sup \mathcal{S}) \}, \quad S^{\prime \prime} \neq S^{\prime} \neq S . \\
\end{array}
\end{equation}

Each event $(E^{\uparrow},S) \in \mathcal{E}^{\prime}$ is the description of an event $(F^{\uparrow}, \sup \mathcal{S}) \in \mathcal{E}$ within the least expressive state-space $S \in \mathcal{S}$. For any events $(E^{\uparrow}, S),(F^{\uparrow}, S^{\prime}) \in \mathcal{E}^{\prime}$, the condition $(E^{\uparrow} \cap \sup \mathcal{S}) = (F^{\uparrow} \cap \sup \mathcal{S})$ implies that both $(E^{\uparrow}, S),(F^{\uparrow}, S^{\prime})$ refer to descriptions of a same event within the non-ordered state-spaces $S,S^{\prime}$, see Eq. \ref{rp1}. Applying Eq. \ref{rp1} in our example in Fig. \ref{hde}, we obtain the full collection of events\footnote{For any lattice $(\mathcal{S}, \succeq)$, the empty event $(\emptyset, \sup \mathcal{S}) \in \mathcal{E}$ is defined as the meet of all events $ \bigwedge_{\mathcal{E}} (E^{\uparrow}, S) = (\emptyset, \sup \mathcal{S}) \in \mathcal{E}$, see Eq. \ref{ip2}. The pair $(\emptyset, \sup \mathcal{S}) \in \mathcal{E}$ is a well-defined event that is not possible to be described within less expressive state-spaces $S \neq \sup \mathcal{S}$, for we include it in the collection $\mathcal{E}^{\prime}$ in Eq. \ref{rp1}.} equal to $\mathcal{E}^{\prime} = \{ (\{a \}^{\uparrow}, S_{a}), (\{ \neg a \}^{\uparrow}, S_{a}), (\{b \}^{\uparrow}, S_{b}), (\{ \neg b \}^{\uparrow}, S_{b}), (\{c_{2} \}^{\uparrow}, S_{c})$, $(\{c_{1}, c_{3} \}^{\uparrow}, S_{c}), (\emptyset, S_{c}), (\Omega, \emptyset) \}$.

For any events $(E^{\uparrow}, S^{\prime}), (F^{\uparrow}, S^{\prime \prime}) \in \mathcal{E}$, define a new partial order as $\sqsubseteq_{S} : (F^{\uparrow}, S^{\prime \prime}) \sqsubseteq_{S} (E^{\uparrow},S^{\prime})$ iff $(F^{\uparrow} \cap S) \subseteq (E^{\uparrow} \cap S)$. The partial order $\sqsubseteq_{S}$ is a conditional set inclusion $\subseteq$ regarding only the states in the state-space $S \in \mathcal{S}$, these states may be included in the upper sets $E^{\uparrow}, F^{\uparrow} \in 2^{\Omega}$. For any event with base-space $S^{\prime}$, filtered state-spaces $S \in \mathcal{S}_{S^{\prime}}$ imply non-empty intersections\footnote{Except for the meet $(\emptyset, \sup \mathcal{S})$.}, i.e. for any $(E^{\uparrow}, S^{\prime}) \in \mathcal{E}, E^{\uparrow} \neq \emptyset$, we have $E^{\uparrow} \cap S \neq \emptyset$ for all $S \in \mathcal{S}_{S^{\prime}}$.

The partial order $\sqsubseteq_{S}$ provides an ordering approach for events in $\mathcal{E}$ which are non-ordered with respect to $\leqslant$. For example, in Fig. \ref{hde}, e.g. the events $(\{a \}^{\uparrow}, S_{a}), (\{ b \}^{\uparrow}, S_{b})$ cannot be ordered with respect to $\leqslant$. Nonetheless, we obtain the order $(\{ b \}^{\uparrow}, S_{b}) \sqsubseteq_{S_{c}} (\{a \}^{\uparrow}, S_{a})$, which follows from the condition $(\{b \}^{\uparrow} \cap S_{c}) = \{ c_{1} \} \subseteq (\{a \}^{\uparrow} \cap S_{c}) = \{ c_{1}, c_{2} \}$.

Denote by $\sqsubseteq := \sqsubseteq_{\sup \mathcal{S}}$ the partial order $\sqsubseteq_{\sup \mathcal{S}}$ with respect to the most expressive state-space $\sup \mathcal{S}$. We obtain a new partially ordered set $(\mathcal{E}^{\prime}, \sqsubseteq)$ bounded by the meet $\inf \mathcal{E}^{\prime} = (\emptyset, \sup \mathcal{S})$ and the join $\sup \mathcal{E}^{\prime} = (\Omega, \emptyset)$, the events satisfying $(E^{\uparrow}, S), (F^{\uparrow}, S^{\prime}) \in \mathcal{E}^{\prime}$, $(F^{\uparrow}, S^{\prime}) \sqsubseteq (E^{\uparrow}, S)$ iff $ (F^{\uparrow} \cap \sup \mathcal{S}) \subseteq (E^{\uparrow} \cap \sup \mathcal{S})$. The new poset $(\mathcal{E}^{\prime}, \sqsubseteq)$ orders all the least expressive descriptions of all events in the full lattice of events $(\mathcal{E}, \leqslant)$, all ordered by decreasing expressiveness of events.

The poset of least expressive events $(\mathcal{E}^{\prime}, \sqsubseteq)$ provides a reduced order-preserving representation of the full lattice of events $(\mathcal{E}, \leqslant)$.

\begin{theorem} \label{t04}
There exists an order preserving $h : \mathcal{E} \rightarrow \mathcal{E}^{\prime}$ between the lattice of events $(\mathcal{E}, \leqslant)$ and the poset of least expressive events $(\mathcal{E}^{\prime}, \sqsubseteq)$.
\begin{proof}
For the posets $(\mathcal{E}, \leqslant), (\mathcal{E}^{\prime}, \sqsubseteq)$, a monotone order preserving $h : \mathcal{E} \rightarrow \mathcal{E}^{\prime}$ is defined as for all $(E^{\uparrow}, S), (F^{\uparrow}, S^{\prime}) \in \mathcal{E}$, if $(F^{\uparrow}, S^{\prime}) \leqslant (E^{\uparrow}, S)$, then $h((F^{\uparrow}, S^{\prime})) \sqsubseteq h((E^{\uparrow}, S))$ \cite{gratzer2011,davey2002}. By the definition of the partial order $\leqslant$, we have the condition $(F^{\uparrow}, S^{\prime}) \leqslant (E^{\uparrow},S) \rightarrow \sup((E^{\uparrow},S), (F^{\uparrow}, S^{\prime})) = (E^{\uparrow},S)$, $\inf((E^{\uparrow},S), (F^{\uparrow}, S^{\prime})) = (F^{\uparrow},S^{\prime})$. By the definition of the partial order $\sqsubseteq$, we have the same condition $(F^{\uparrow}, S^{\prime}) \sqsubseteq (E^{\uparrow},S) \rightarrow \sup((E^{\uparrow},S), (F^{\uparrow}, S^{\prime})) = (E^{\uparrow},S)$, $\inf((E^{\uparrow},S), (F^{\uparrow}, S^{\prime})) = (F^{\uparrow},S^{\prime})$.
\end{proof}
\end{theorem}

Refer to $|S| : S \rightarrow \mathbb{N}$ as the cardinality of $S$. For all $S, S^{\prime} \in \mathcal{S}$, if more expressive state-spaces have higher cardinality\footnote{Recall that projections $r^{S^{\prime}}_{S} : S^{\prime} \rightarrow S$ are surjective by definition. Studies usually assume that more expressive state-spaces $S^{\prime} \succeq S$ include more states, see \cite{fukuda2024,heifetz2006,heifetz2013,schipper2021}.}, $S^{\prime} \succeq S$, $|S^{\prime}| - |S| \neq 0$, i.e. space $S^{\prime}$ is refined, including more states than $S$, then the poset of least expressive events $(\mathcal{E}^{\prime}, \sqsubseteq)$ provides sufficient information to recover the original lattice of events $(\mathcal{E}, \leqslant)$ from which it derives. In this sense, we say that the poset $(\mathcal{E}^{\prime}, \sqsubseteq)$ provides the same informational content as the complete lattice of events $(\mathcal{E}, \leqslant)$, for this is a main result of our analysis.

\begin{theorem} \label{t06}
Assume that more expressive state-spaces imply higher cardinality, for all $S, S^{\prime} \in \mathcal{S}$, $S^{\prime} \succeq S$, $|S^{\prime}| - |S| \neq 0$. Then, the poset of least expressive events $(\mathcal{E}^{\prime}, \sqsubseteq)$ provides the same informational content as the complete lattice of events $(\mathcal{E}, \leqslant)$.
\begin{proof}
Let the collection of events $\mathcal{E}^{\prime} \subseteq \mathcal{E}$ be defined in Eq. \ref{rp1}. The empty event $(\emptyset, \sup \mathcal{S}) \in \mathcal{E}^{\prime}$ is always an element of $\mathcal{E}^{\prime}$, and it provides information about the states in the most expressive state-space $\sup \mathcal{S}$. For every event $(E^{\uparrow}, S) \in \mathcal{E}^{\prime}$, the base-space $S$ provides information about the states in the least expressive state-space $S$ which describes it. For each non-empty event $(E^{\uparrow}, S) \in \mathcal{E}^{\prime}$, the upper set $E^{\uparrow}$ provides information about states in the filtered state-spaces $S^{\prime} \in \mathcal{S}_{S}$. Any order $(\emptyset, \sup \mathcal{S}) \sqsubseteq (E^{\uparrow}, S)$ in the poset $(\mathcal{E}^{\prime}, \sqsubseteq)$ implies an order $(\emptyset, \sup \mathcal{S}) \leqslant (F^{\uparrow}, S^{\prime}) \leqslant (E^{\uparrow}, S)$ in the lattice $(\mathcal{E}, \leqslant)$ for states satisfying $\omega \in E^{\uparrow}, \omega \in S^{\prime}$. Since more expressiveness implies higher cardinality, for all $S, S^{\prime} \in \mathcal{S}$, $S^{\prime} \succeq S$, $|S^{\prime}| - |S| \neq 0$, and since projections across state-spaces $r^{S^{\prime}}_{S}$ are surjective by definition, therefore every state-space $S \in \mathcal{S}$ is a base-space of at least one event $(E^{\uparrow}, S) \in \mathcal{E}^{\prime}$. 
\end{proof}
\end{theorem}

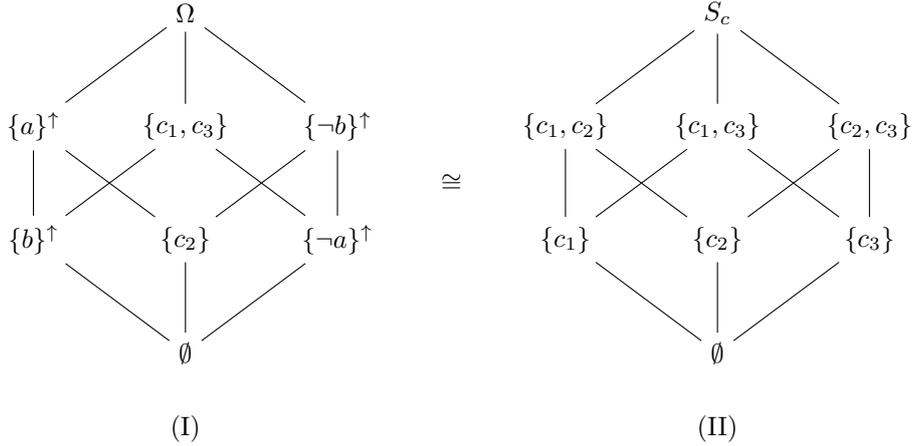
\begin{figure}[htb!]
\begin{center}
\begin{tikzpicture}
	\node (min) at (0,1) {$ \emptyset$};
	\node (c1) at (-2,2.5) {$ \{ b \}^{\uparrow}$};
	\node (c2) at (0,2.5) {$ \{ c_{2} \}$};
	\node (c3) at (2,2.5) {$ \{ \neg a \}^{\uparrow}$};
	\node (cc1) at (-2,4) {$ \{ a \}^{\uparrow}$};
	\node (cc2) at (0,4) {$ \{ c_{1}, c_{3} \}$};
	\node (cc3) at (2,4) {$ \{ \neg b \}^{\uparrow}$};
	\node (max) at (0,5.5) {$ \Omega$};
	\node (mmin) at (0,0) {(I)};
	\node (iso) at (3.5,3.25) {$\cong$};
	\node (Cmin) at (7,1) {$ \emptyset$};
	\node (Cc1) at (5,2.5) {$ \{ c_{1} \}$};
	\node (Cc2) at (7,2.5) {$ \{ c_{2} \}$};
	\node (Cc3) at (9,2.5) {$ \{ c_{3} \}$};
	\node (Ccc1) at (5,4) {$ \{ c_{1}, c_{2} \}$};
	\node (Ccc2) at (7,4) {$ \{ c_{1}, c_{3} \}$};
	\node (Ccc3) at (9,4) {$ \{ c_{2}, c_{3} \}$};
	\node (Cmax) at (7,5.5) {$ S_{c} $};
	\node (Cmmin) at (7,0) {(II)};
	\draw (min) -- (c1) -- (cc1) -- (max);
	\draw (min) -- (c2) -- (cc1);
	\draw (c1) -- (cc2) -- (max);
	\draw (c2) -- (cc3) -- (max);
	\draw (min) -- (c3) -- (cc2);
	\draw (c3) -- (cc3);
	\draw (Cmin) -- (Cc1) -- (Ccc1) -- (Cmax);
	\draw (Cmin) -- (Cc2) -- (Ccc1);
	\draw (Cc1) -- (Ccc2) -- (Cmax);
	\draw (Cc2) -- (Ccc3) -- (Cmax);
	\draw (Cmin) -- (Cc3) -- (Ccc2);
	\draw (Cc3) -- (Ccc3);
\end{tikzpicture}
\end{center}
\caption{(I) Hasse diagram for the poset of the least expressive events $(\mathcal{E}^{\prime}, \sqsubseteq)$ deriving from the lattice of events in Fig. \ref{hde}, with $\mathcal{E}^{\prime} = \{ (\{a \}^{\uparrow}, S_{a}), (\{ \neg a \}^{\uparrow}, S_{a})$, $(\{b \}^{\uparrow}, S_{b}), (\{ \neg b \}^{\uparrow}, S_{b}), (\{c_{2} \}^{\uparrow}, S_{c}), (\{c_{1}, c_{3} \}^{\uparrow}, S_{c}), (\emptyset, S_{c}), (\Omega, \emptyset) \}$. (II) Hasse diagram for the lattice of set inclusion $(2^{S_{c}}, \subseteq)$, $S_{c} = \{ c_{1}, c_{2}, c_{3} \}$. We omit the base-spaces of events by applying a concise notation equal to $(E^{\uparrow},S) = E^{\uparrow}$ for all events $(E^{\uparrow},S) \in \mathcal{E}^{\prime}$, the least-expressiveness condition $E \subseteq S$ implying unambiguous base-spaces.} \label{hdp}
\end{figure}

For example, the lattice of events $(\mathcal{E}, \leqslant)$ in Fig. \ref{hde} derives the poset of least expressive events $(\mathcal{E}^{\prime}, \sqsubseteq)$ in Fig. \ref{hdp} (we apply the same simplified notation as in Fig. \ref{hde}). We have the orders of expressiveness $S_{c} \succeq S_{a}, S_{c} \succeq S_{b}$, cardinalities $|S_{c}| = 3, |S_{a}| = |S_{b}| = 2$. Theorem \ref{t06} shows that the poset $(\mathcal{E}^{\prime}, \sqsubseteq)$ in Fig. \ref{hdp} provides the same informational content as the complete lattice of events $(\mathcal{E}, \leqslant)$ in Fig. \ref{hde}.

The poset of least expressive events $(\mathcal{E}^{\prime}, \sqsubseteq)$ may include descriptions of a same event within non-ordered state-spaces, i.e. events $(E^{\uparrow}, S),(F^{\uparrow}, S^{\prime}) \in \mathcal{E}^{\prime}$ satisfying $(E^{\uparrow} \cap \sup \mathcal{S}) = (F^{\uparrow} \cap \sup \mathcal{S})$, in which case $(\mathcal{E}^{\prime}, \sqsubseteq)$ is not a lattice. The poset $(\mathcal{E}^{\prime}, \sqsubseteq)$ is a lattice iff we have the condition $(E^{\uparrow} \cap \sup \mathcal{S}) \neq (F^{\uparrow} \cap \sup \mathcal{S})$ for all events $(E^{\uparrow}, S),(F^{\uparrow}, S^{\prime}) \in \mathcal{E}^{\prime}$. Our example in Fig. \ref{hdp} satisfies this condition, therefore it is a lattice.

\begin{theorem} \label{t03}
Assume the condition $(E^{\uparrow} \cap \sup \mathcal{S}) \neq (F^{\uparrow} \cap \sup \mathcal{S})$ for all events $(E^{\uparrow}, S),(F^{\uparrow}, S^{\prime}) \in \mathcal{E}^{\prime}$. Then, the poset of least expressive events $(\mathcal{E}^{\prime}, \sqsubseteq)$ is isomorphic to the lattice of set inclusion $(2^{(\sup \mathcal{S})}, \subseteq)$.
\begin{proof}
An order isomorphism is a bijection referred by $g^{-1} : \mathcal{E}^{\prime} \rightarrow 2^{(\sup \mathcal{S})}$ defined as for all events $(E^{\uparrow}, S), (F^{\uparrow},S^{\prime}) \in \mathcal{E}^{\prime}$, we have $(F^{\uparrow},S^{\prime}) \sqsubseteq (E^{\uparrow}, S)$ iff $g^{-1}((F^{\uparrow},S^{\prime})) \subseteq g^{-1}((E^{\uparrow}, S))$ \cite{gratzer2011,davey2002}. The condition $(E^{\uparrow} \cap \sup \mathcal{S}) \neq (F^{\uparrow} \cap \sup \mathcal{S})$ for all $(E^{\uparrow}, S), (F^{\uparrow},S^{\prime}) \in \mathcal{E}^{\prime}$ implies a bijection between $\mathcal{E}^{\prime}$ and $2^{(\sup \mathcal{S})}$, for we have $(E^{\uparrow} \cap \sup \mathcal{S}) \in 2^{(\sup \mathcal{S})}$ for each event $(E^{\uparrow}, S) \in \mathcal{E}^{\prime}$. By definition, the partial order $\sqsubseteq$ refers to the standard set inclusion $\subseteq$ with respect to the state-space $\sup \mathcal{S}$.
\end{proof}
\end{theorem}

Denote the isomorphism in Theorem \ref{t03} by $(\mathcal{E}^{\prime}, \sqsubseteq) \cong (2^{(\sup \mathcal{S})}, \subseteq)$, e.g. see Fig. \ref{hdp}. It provides the conditions to embed the poset of least expressive events $(\mathcal{E}^{\prime}, \sqsubseteq)$ into the lattice of events $(\mathcal{E}, \leqslant)$. Apply isomorphism $g$, embedding $f$ and antisymmetry\footnote{For any events $(E^{\uparrow},S), (F^{\uparrow}, S^{\prime}) \in \mathcal{E}$, the property of antisymmetry is defined as $(E^{\uparrow},S) \leqslant (F^{\uparrow}, S^{\prime})$ and $(F^{\uparrow}, S^{\prime}) \leqslant (E^{\uparrow},S)$ imply $(E^{\uparrow},S) = (F^{\uparrow}, S^{\prime})$ \cite{gratzer2011,davey2002}.}, and refer to the image of the composite mapping equal to $(f \circ g) (2^{(\sup \mathcal{S})}) = f(\mathcal{E}^{\prime})$, see Theorem \ref{t03}. We find $(\mathcal{E}^{\prime}, \sqsubseteq) \cong (f(\mathcal{E}^{\prime}), \leqslant)$.

\begin{theorem} \label{t05}
Refer to $|\mathcal{E}^{\prime}|$ as the cardinality of $\mathcal{E}^{\prime}$. The following results are equivalent:

1. $(\mathcal{E}^{\prime}, \sqsubseteq) \cong (f(\mathcal{E}^{\prime}), \leqslant)$;

2. $|\mathcal{E}^{\prime}| = |2^{(\sup \mathcal{S})}|$; 

3. $(E^{\uparrow} \cap \sup \mathcal{S}) \neq (F^{\uparrow} \cap \sup \mathcal{S})$ for all events $(E^{\uparrow}, S),(F^{\uparrow}, S^{\prime}) \in \mathcal{E}^{\prime}$.

\end{theorem}

The cardinality condition $|\mathcal{E}^{\prime}| = |2^{(\sup \mathcal{S})}|$ in Theorem \ref{t05} is straightforward and it implies the isomorphism $(\mathcal{E}^{\prime}, \sqsubseteq) \cong (f(\mathcal{E}^{\prime}), \leqslant)$ necessarily, see Theorem \ref{t03}.


\section{Related Literature and Discussion} \label{Related Literature and Discussion}

The lattice model of information structures is first proposed by \cite{heifetz2006}, for it becomes the baseline structure underlying current set-theoretic models of knowledge of agents, e.g. \cite{fukuda2023,fukuda2024,schipper2021,heifetz2013}. For each event defined as a pair of upper sets and base-spaces, see Eq. \ref{ip1}, the knowledge of each agent derives from a generalised operator defined over the collection of all events. It formalises knowledge as an event itself, therefore providing a model of introspective reasoning and mutual information \cite{fukuda2023,fukuda2024,heifetz2006,heifetz2013,schipper2021}.

The lattice of events based on the order of expressiveness $\leqslant$ derives some well-defined properties applicable to the modelling of rational knowledge, e.g. monotonicity, positive introspection, and the property of truth \cite{fukuda2024,heifetz2013,schipper2021}. The lattice of events also proposes a solution to major inconsistencies found in the standard state-space model, notably regarding the representation of unawareness of agents \cite{modica1994,modica1999,dekel1998}.

One relevant application of the lattice model refers to the analysis of the special conditions under which speculative trade occurs. Classical "no-trade" theorems in economics and finance demonstrate that trade is not feasible when agents have mutual understanding of their knowledge structures under the standard partitional model, for rational agents become indifferent to trade after reasoning about each other's knowledge \cite{aumann1976,milgrom1982,tirole1982,rubinstein1998}. Studies show that speculative trade requires agents to be unaware of some payoff-relevant states of the world \cite{fukuda2023,fukuda2024,heifetz2006,heifetz2013,schipper2021}, for the lattice of events provides an adequate model for analysis.

We derive a reduced poset of least expressive events which provides the same informational content as the original lattice of events which derives it, under the baseline model. It implies that agents can recover the complete lattice of events by means of the least expressive events included in our poset only. In models where agents share information or reason about each other's knowledge, the poset of least expressive events may be a more efficient representation of the complete information structure of agents since it carries full information in a subcollection of the original collection of all events.

We show the conditions on how to embed the poset of least expressive events into the lattice of events, so that the orders of expressiveness defined within each information structure provide isomorphic models.

Our analysis derive from a conditional subset inclusion regarding the states defined within a specific state-space, for it provides the ordering of events based on specific levels of expressiveness. It may apply to order events that are defined with respect to generalisations of the standard lattice model, where the orders of expressiveness of state-spaces do not derive from conventional space refinement.



\end{document}